\documentclass[5p, twocolumn]{elsarticle}  

\usepackage{graphicx}
\usepackage{amsmath}
\usepackage{amssymb}
\usepackage{hyperref}
\usepackage{natbib}

\hypersetup{
colorlinks,
citecolor=blue,
filecolor=blue,
linkcolor=blue,
urlcolor=blue }

\newcommand{\be}{\begin{equation}}
\newcommand{\ee}{\end{equation}}
\newcommand{\bse}{\begin{subequations}}
\newcommand{\ese}{\end{subequations}}
\newcommand{\bea}{\begin{eqnarray}}
\newcommand{\eea}{\end{eqnarray}}
\newcommand{\ba}{\begin{array}}
\newcommand{\ea}{\end{array}}

\begin{document}	

\title{ Scalable Multi-node Fast Fourier Transform on GPUs}


\author[iitk]{Manthan Verma}
\author[iitk]{Soumyadeep Chatterjee}
\author[cdac_pune]{Gaurav Garg}
\author[cdac_banglore]{Bharatkumar Sharma}
\author[iitke]{Nishant Arya}
\author[iitk]{Shashi Kumar}
\author[iitkm]{Anish Saxena}
\author[iitk]{Mahendra K.Verma \corref{cor1}}
\cortext[cor1]{Corresponding author}
\address[iitk]{Department of Physics, Indian Institute of Technology Kanpur, Kanpur 208016, India}
\address[iitkm]{Department of Mechanical Engineering, Indian Institute of Technology Kanpur, Kanpur 208016, India}
\address[iitke]{Department of Electrical Engineering, Indian Institute of Technology Kanpur, Kanpur 208016, India}
\address[cdac_pune]{NVIDIA Graphics Pvt Ltd, Pune 411006, India}
\address[cdac_banglore]{NVIDIA Graphics Pvt Ltd, Bangalore 560045, India}

\begin{abstract}
In this paper, we present the details of our multi-node GPU-FFT library, as well its scaling on {\em Selene} HPC system. Our library employs slab decomposition for data division and MPI for communication among GPUs. We performed GPU-FFT on $1024^3$, $2048^3$, and $4096^3$ grids using a maximum of 512 A100 GPUs. We observed good scaling for $4096^3$ grid with 64 to 512 GPUs. We report that the timings of multicore FFT of $1536^3$ grid with 196608 cores of Cray XC40 is comparable to that of  GPU-FFT of $2048^3$ grid with 128 GPUs. The efficiency of GPU-FFT is due to the fast computation capabilities of A100  card and efficient communication via NVlink.
\end{abstract}

\maketitle

\section{Introduction}
\label{sec:intro}

Parallel Fast Fourier Transform (FFT) is an important application of signal processing and spectral solvers~\cite{Boyd:book:Spectral}. The leading parallel FFT libraries available today are FFTW, P3DFFT, PFFT, cuFFTXT, etc. Most of these libraries are for multicore systems, and they have been scaled reasonably well up to 500000 processors. In this paper we present the results of our new multi-node GPU-FFT.

Cooley and Tukey~\cite{Cooley:AMS1965} invented  FFT to speed up Fourier transform of $N^3$ data from $O(N^6)$ to $O(N^3 \log N^3)$. FFT has been parallelized for multicore and multiGPU systems. One of the most prominent FFT libraries is FFTW (Fastest Fourier Transform in the West)~\cite{FFTW:web,Frigo:IEEE2005}. Many spectral applications have adopted this library.  However, FFTW is based on slab decomposition, hence it is not suitable when the number of available processors is more than the number of grid points along an axis ($N$ of $N^3$ grid).

Pencil-based parallel FFTs can  make use of large number of processors available in modern supercomputers. A pencil-based FFT can employ a maximum of $N^2$ processors for a $N^3$ grid. The leading pencil-based FFTs are P3DFFT~\cite{Pekurovsky:SIAM2012}, FFTK~\cite{Chatterjee:JPDC2018, samar:2021}, and PFFT~\cite{Pippig:SIAM2013}. For P3DFFT, Pekurovsky~\cite{Pekurovsky:SIAM2012} reported that the computation time scales as $p^{-1}$, while the  communication time scales as $p^{-2/3}$, where $p$ is the number of compute cores.  This scaling was derived based on runs for a $8192^3$  grid on Cray XT5 having  65536 cores and a 3D torus interconnect. Chatterjee {\em et al.}~\cite{Chatterjee:JPDC2018} ran FFTK library on {\em Shaheen I} (Blue Gene/P) and {\em Shaheen II} (Cray XC40) for grids up to $8192^3$.  They observed good scaling (approximately $p^{-0.7}$) up to 196608 compute cores of Cray XC40. Pippig~\cite{Pippig:SIAM2013} created the PFFT library that exhibits a similar scaling. Mininni {\em et al.}~\cite{Mininni:PC2011} employed hybrid scheme (MPI + OPENMP) for their FFT that scales well for  $1536^3$ and  $3072^3$ grids using 20000 cores with 6 and 12 threads on each socket.

Multicore systems are very much in vogue today.  However, at present, many HPC systems are employing GPUs as accelerators~\cite{GPU_strategies:2013jpdc,efficiacy_gpu_mpi_scientific:2013,cpu_gpu_comm_latency_fine_grained:2013,cpu_gpu_matrix:2017}. The peak single-precision and double-precision performance of A100 GPU are 19.5 TFLOPS and 9.7 TFLOPS~\cite{A100_sheet:web} respectively, that far exceeds the peak performance of 2 TFLOPS (double precision) of Rome processor.  The newest A100 card has 80 GB VRAM that enables execution of large HPC applications in a GPU cluster. Inside a DGX box,  NVlink  provides very fast communication among its 8 GPU cards. The bandwith of NVlink is 600 GB/sec~\cite{nvlinks_docs:web}, which is 10X the bandwidth of PCIe Gen 4. Due to these superlative performance of GPUs, many powerful HPC systems provide these GPUs as accelerator. As on November 2021, GPU cluster {\em Selene}, which is the sixth fastest HPC system in the world \cite{top_500highlights:web}, contains 540 DGX boxes with a total 4320 A100 cards.  Due to the availability of  such efficient computing systems, many multicore  applications have been ported to GPUs. We also remark that GPUs are preferred hardware for artificial intelligence (AI) applications.

Regarding GPU-FFT, at first, NVIDIA provided a single-GPU FFT library called  cuFFT. Later, a new library called cuFFTXT~\cite{cufft_docs:web} was provided that supports FFT on the multiple GPUs of a single node. The other GPU based FFTs are DiGPUFFT~\cite{Czechowski:CP2012}, heFFTe~\cite{heffte_2020,hefftvsmpi_cpu_gpu_v100}, AccFFT~\cite{acfft:2015}, cusFFT~\cite{cusFFTAH:Wang2016}, etc. In a recent work, Ravikumar {\em et al.}~\cite{young_PK:ACM2019} constructed a GPU-based FFT and a pseudo-spectral code for turbulence simulations and ran it on {\em Summit}. They employed 3072 nodes of {\em Summit} (containing 6 V100 GPUs per node) to perform FFT of grids up to $18432^3$.  They observed a GPU to CPU speedup of 4.7 for $12288^3$ grid and a speedup of 2.9 for $18432^3$ grid.  Recently, Bak {\em et al.}~\cite{Bak:PC2022} measured the performance of a synchronous non-batched version of this GPU-FFT on 1024 nodes of {\em Summit} and  obtained a maximum GPU to CPU speedup of $2.57$ for $12228^3$ grid.

 We have created a new CUDA-based multi-node GPU-FFT, which is available for download at github\footnote{\url{https://github.com/Manthan-Verma/GPU_FFT.git}}. For data division, we employ slab decomposition because the number of available GPUs is typically much smaller than the grid size along any direction.  Our library uses MPI functions for communication among the GPUs within and outside the compute nodes. Also, we employ  cuFFT for the  one-dimensional and two-dimensional FFTs within a GPU. We optimized our library on the {\em Selene} cluster and ran it for $1024^3$, $2048^3$, and $4096^3$ grids using a maximum of 512 GPU cards (or 64 nodes). Our FFT library scales well for large grids with proportionally large number of GPUs. In this paper, we present the scaling results of our FFT library. We remark that recently, NVIDIA has announced a new multi-node FFT library named cuFFTMp~\cite{cufftmp:web}. However, cuFFTMp is not  available for public use at present. 

Like other HPC applications, GPU-FFT too provides significant speedup in comparison to multicore performance~\cite{young_PK:ACM2019}. We show that 128 A100 cards yield  performance comparable to 196608 cores of Cray XC40.  Such a speedup encourages us to optimize the GPU-FFT even further. 

FFT is a backbone for  spectral solvers.  Researchers have performed high-resolution turbulence simulations using FFT~\cite{Yokokawa:CP2002,Kaneda:PF2003,Donzis:2008us,Yeung:PNAS2015,Rosenberg:PF2015,Donzis:PF2008,Donzis:FTC2010,Donzis:JFM2010Bottleneck,Yeung:PF2005,Yeung:JFM2013,Rorai:PRE2015,Dallas:PRL2015}.  Yokokawa {\em et al.}~\cite{Yokokawa:CP2002} performed one of the first high-resolution turbulence simulations on $4096^3$ grid using the {\em Earth Simulator}. At present, the largest turbulent flow simulated is on a $18432^3$ grid~\cite{young_PK:ACM2019, Yeung:PRF2020}; this application employs GPUs. Parallel FFT is employed in density functional theory, and as well as in signal processing of images and speech. Radio astronomers too employ parallel FFT for data analysis. Due to lack of space, we do not list the details of these applications.

An outline of our paper is as follows:   Section~\ref{sec:numerical_scheme} contains an introduction to parallel FFT and data decomposition. In Section~\ref{sec:parallel}, we describe the implementation of parallel FFT in GPUs. In  Section~\ref{sec:results:main}, we report the scaling results of our GPU-FFT, while in Section~\ref{sec:comparisons}, we compare the results of our library with other FFT libraries. We conclude in Section~\ref{sec:conclusions}.

\section{Introduction to parallel  FFT and data decomposition}
\label{sec:numerical_scheme}

The forward and inverse three-dimensional (3D) discrete FFTs are defined as follows:
\bea
\hat{f}(k_x, k_y, k_z) & = & \sum_{j_x} \sum_{j_y} \sum_{j_z}  
 f(j_x, j_y, j_z) \times \nonumber \\
 && \exp\left( -2\pi i \left[\frac{j_x k_x}{N_x}+\frac{j_y k_y }{N_y}+\frac{j_z k_z}{N_z}\right] \right),
\label{eq:forward} \nonumber \\ \\
f(j_x, j_y, j_z) & = & \sum_{k_z}  \sum_{k_y}  \sum_{k_x} 
\hat{f}(k_x, k_y, k_z) \times \nonumber \\
&& \exp\left( 2\pi i \left[\frac{j_x k_x}{N_x}+\frac{j_y k_y }{N_y}+\frac{j_z k_z}{N_z}\right] \right) .
\label{eq:inverse} \nonumber \\
\eea
Here, $N_x, N_y, N_z$ are the number of grid points along $x,y,z$ directions respectively, $f(j_x, j_y, j_z)$ is the field variable at real-space grid point $(j_x, j_y, j_z)$, and $\hat{f}(k_x, k_y, k_z)$ is the Fourier transform at the wavenumber $(k_x, k_y, k_z)$. Note that in real space, $f(x,y,z)$ is periodic in a box of size $(2\pi)^3$ with $N_x  N_y N_z$ grid points.  In this paper, we take the field $f(x,y,z)$ to be real. Hence,
\be
\hat{f}(-k_x, -k_y, -k_z) = \hat{f}^*(k_x, k_y, k_z).
\ee
Therefore, we store only half of the $\hat{f}$ array. For example, the range of wavenumbers could be 
\be
k_x = 0:N_x-1;~~~k_y = 0:N_y-1;~~~k_z = 0:N_z/2+1.
\ee
Note, however, that the direction of wavenumber reduction could be along any direction. In this paper, we follow the index notation of FFTW~\cite{Frigo:IEEE2005}. For simplification, in this paper, we take $N_x = N_y = N_z = N$.

A trivial implementation of Eqs.~(\ref{eq:forward},\ref{eq:inverse}) would take $O(N^6)$ floating point operations. However, a clever scheme by Cooley and Tukey~\cite{Cooley:AMS1965} takes  $O(N^3 \log(N^3))$ floating point operations. The latter implementation is called Fast Fourier Transform (FFT). We will discuss its parallel GPU implementation in the next section.

We can perform 3D FFT on a single processor for a small grid, say $128^3$. But, we need many processors or GPUs for performing FFT on large grids. For parallel computation, the data is divided in two ways: 
\begin{enumerate}
    \item \textit{Slab decomposition}: In this decomposition, the data is divided into slab. For $p$ processors, we could divide data along the $x$ axis with each processor/GPU containing $(N_x/p) \times N_y \times N_z$ real-space data points. Here, the maximum number of processors we can employ is $N_x$.  Refer to Frigo and Johnson~\cite{Frigo:IEEE2005} for further details.      In the following section we will show how FFT is performed for this division.  
    
    \item \textit{Pencil decomposition}: In this decomposition, the data is divided into columns. For example, the total number of processors $p$ could be arranged on a grid as $p_x \times p_y = p$, with each processor containing $(N_x/p_x) \times (N_y/p_y) \times N_z$ real-space data. Here, the maximum number of processors we can employ is $N_x \times N_y$, which is far larger than what is possible with slab decomposition. Refer to Pekurovsky~\cite{Pekurovsky:SIAM2012} and Chatterjee {\em et al.}~\cite{Chatterjee:JPDC2018} for details on pencil decomposition. 
\end{enumerate}

Large number of cores are available in modern multicore HPC systems. Hence, pencil decomposition is preferred for such  systems. However,  it is optimum to employ slab decomposition when we have small number of processors or GPU. In this paper, we work with a maximum of 512 GPUs to perform FFTs on grids of the sizes $1024^3$, $2048^3$, and $4096^3$.  The slab decomposition is preferred for  these configurations. In the next section, we will describe an implementation of GPU-FFT.


\begin{figure*}[htbp]
\begin{center}
\includegraphics[scale=0.47]{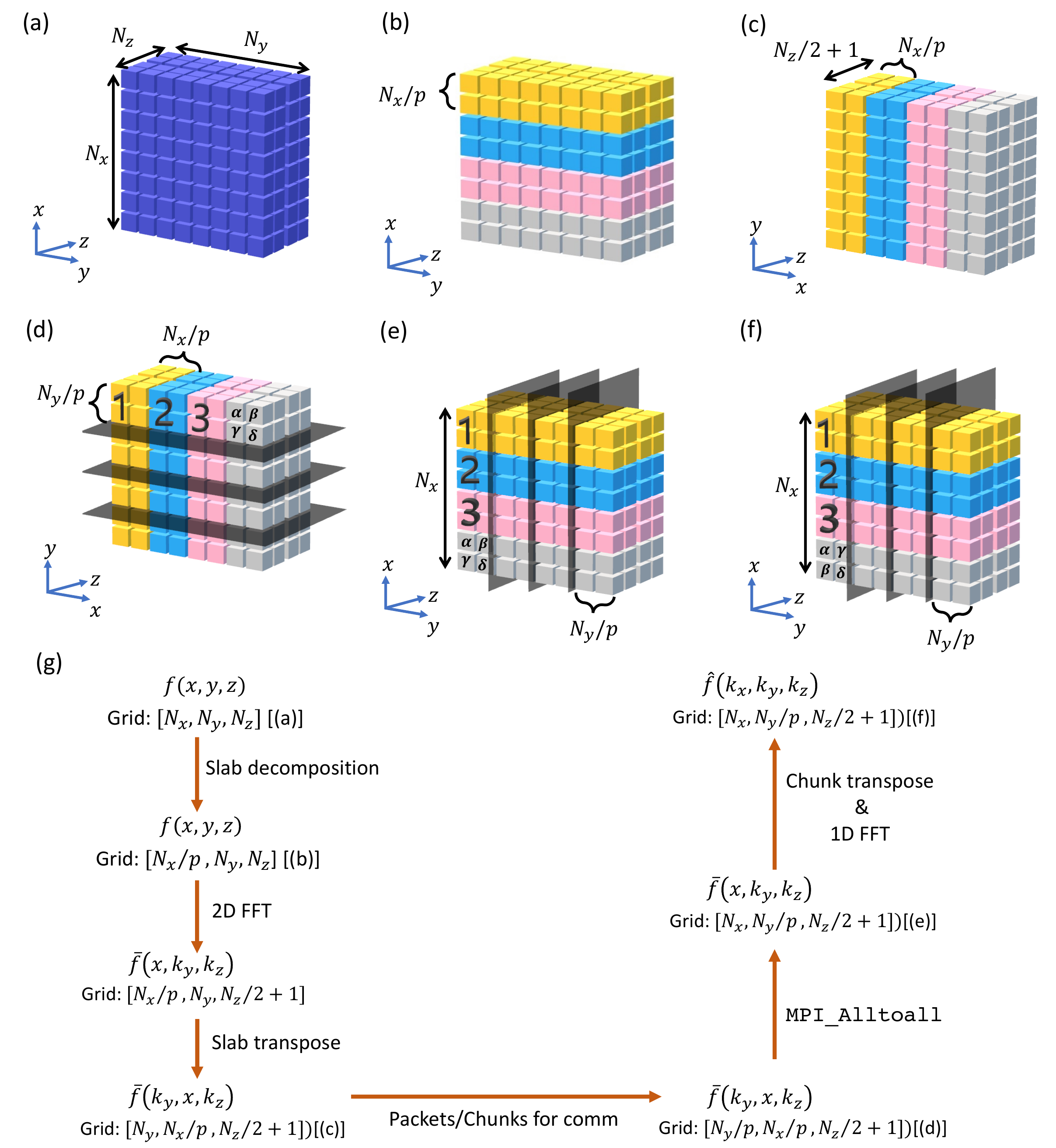}
\end{center}
\setlength{\abovecaptionskip}{0pt}
\caption{(a) Initial 3D real-space data in the CPU. (b) Data divided into $p$ (here 4) segments and sent to  $p$ GPUs. Each GPU  contains $N_{x}/p \times N_{y} \times N_{z}$ data points. 2D Forward FFT is performed on each plane of the data by the respective GPUs.  (c) The $xy$ transpose is performed on the local data of (b) that results into configuration of Fig. (c).  (d,e) We perform \texttt{Alltoall} communication on the slab  that leads to configuration of Fig.\ref{fig:slab_decomposition} (e). Note that after the communication, the data along $x$ axis is not in proper order [see $(\alpha,\beta, \gamma,\delta)$ segment]. We perform local transpose on such data segments, called \textit{chunks}.
(f) Configuration after the local transpose on chunks. The data along $x$ are now in proper order. 1D C2C  cuFFT is performed along the $x$ direction for all the columns of the $xz$ slabs. (g) A schematic diagram of the algorithm. }
\label{fig:slab_decomposition}
\end{figure*}

\section{Parallel FFT implementation on GPUs}
\label{sec:parallel}


As described above, it is preferable to employ slab decompostion for GPUs.  Also, CUDA implementation of slab decomposition is  simpler than that of  pencil decomposition. The algorithm adopted by us is described below. We illustrate the steps involved in Fig.~\ref{fig:slab_decomposition}. 

The real-space field, depicted in Fig.~\ref{fig:slab_decomposition}(a), resides in  CPU's memory. Suppose we have $p$ GPUs available for the FFT operation. Then, CPU divides the data into $p$ slabs along the $x$ axis and then performs \texttt{cudaMemcpy} to the $p$ GPUs, as illustrated in Fig.~\ref{fig:slab_decomposition}(b) for $p=4$. Each GPU now contains $N_{x}/p \times N_{y} \times (N_{z} / 2 + 1)$ data points. We start from here and state the steps of  the forward FFT. Note that $z$ is the fastest direction of the array indices, and $x$ is the slowest direction.

\begin{enumerate}
    \item \textit{FFT along} $yz$:  Each GPU performs forward 2D  real-to-complex (R2C) FFT  using \texttt{cufft} on all the $yz$ planes of the slab  of Fig.~\ref{fig:slab_decomposition}(b). These operations are performed on all the planes in the slab  in batch mode with $N_{x}/p$ number of batches.
    
    \item \textit{Local $xy$ transpose}: We perform local $xy$ transpose within each GPU. The resulting data is as shown in Fig.~\ref{fig:slab_decomposition}(c). After the transpose, the data is arranged as $N_{y} \times N_{x}/p \times (N_{z} / 2 + 1)$. 
    
    \item \texttt{Alltoall} \textit{communication}: Next, we perform  \texttt{Alltoall} communication to bring  the data along $x$ direction in  GPUs.  We observe that the present system implementation of \texttt{MPI\_Alltoall} collective call takes a long time due to unnecessary copying of data. Hence,   we employ a combination of asynchronous \texttt{MPI\_Isend} and \texttt{MPI\_Irecv}, as well as  \texttt{MPI\_Waitall} and \texttt{cudaMemcpy}, instead of  \texttt{MPI\_Alltoall} collective communication. Figure~\ref{fig:slab_decomposition}(d,e) illustrate the data  before and after the \texttt{Alltoall} communication.

\hspace{0.25cm} For \texttt{MPI\_Isend} and \texttt{MPI\_Irecv} operations, we create chunks of data. An example of chunk is shown as $(\alpha, \beta, \gamma, \delta)$ in Fig.~\ref{fig:slab_decomposition}(d,e). The size of each chunk is $N_{x}/p \times N_{y}/p \times (N_{z} / 2 + 1)$, and the symbols $1, 2, 3$ in Fig.~\ref{fig:slab_decomposition}(d,e,f) specify the order of different chunks in  GPUs.  The sequence of  data segment $(\alpha, \beta, \gamma, \delta)$ in the chunk remains unchanged during the \texttt{Alltoall} operation, as shown in Fig. \ref{fig:slab_decomposition}(d,e).

    \item \textit{Local $xy$ transpose of the chunks}: After  \texttt{Alltoall} operation, the data along the $x$ axis is not in proper sequence for a FFT operation. Note that the first row along the $x$ axis need to have $\beta$ after $\alpha$ for 1D C2C FFT.  A proper sequencing of the data is achieved by making a local transpose of the chunk data along $xy$.  The final data after the local chunk transpose is shown in Fig. \ref{fig:slab_decomposition}(f).   
    
    \item \textit{FFT along $x$}: Lastly, we perform 1D strided and batched complex-to-complex (C2C) FFT using \texttt{cufft} along the $x$ axis. We employ $N_{y}/p\times (N_{z}/2+1)$ batches for the operation. The stride of $N_{y}/p \times (N_{z} / 2 + 1)$ is used because the data along the $x$ axis are not consecutive. 
    
\end{enumerate}

A schematic diagram of our algorithm is depicted in Fig.~\ref{fig:slab_decomposition}(g). In addition, we incorporate the following features in our library for optimization:
\begin{enumerate}
    
    \item For real-to-complex, complex-to-complex, and complex-to-real Fourier transforms, we use the same work-space pointer  that leads to smaller memory requirements.
    
    \item Using GPUDirect RDMA and CUDA-aware MPI~\cite{comm_GPU_acc_thesis:2018}, we avoid unnecessary data transfers between device and host. Note that copying data between the host and the device is  expensive. We remark that our strategy differs from that of Ravikumar {\em et al.}~\cite{young_PK:ACM2019}, who employ device-host data transfer for \texttt{Alltoall} operation. We believe that device-to-device communication should yield fast communication.
    
    \item Self communication during a \texttt{MPI\_Alltoall} collective call is  relatively slow due to large number of  device to device copying~\cite{ucx_bug:web}. We overcome this bottleneck by using \texttt{MPI\_Isend}, \texttt{MPI\_Irecv},  \texttt{MPI\_Waitall} and \texttt{cudaMemcpy} for \texttt{Alltoall} operation.
    
    \item Chatterjee {\em et al.}~\cite{Chatterjee:JPDC2018} employed \texttt{MPI\_Vector} during \texttt{MPI\_Alltoall} in order to avoid the local transpose; this feature leads to some gain the performance. We tested the above in our CUDA implementation. Unfortunately, such arrangements do not yield performance improvements for  GPUs.
    
\end{enumerate}

For the inverse FFT, we employ the above steps in a reverse order.  In the next section, we report the scaling results of our GPU-FFT on {\em Selene}.

\section{Performance analysis of GPU-FFT}
\label{sec:results:main}

In this section, we analyze the efficiency of our GPU-FFT. We perform our runs on  NVIDIA's {\em Selene} supercomputer for both single and double precision. {\em Selene}, which is the sixth ranked on November 2021's top500 list \cite{top_500highlights:web}, has 540 DGX boxes (or nodes) and 1120 TB of system memory.  Each  DGX box contains two Rome processors and 8 A100 GPU cards (each with 80 GB  VRAM). The GPUs inside the DGX box communicate to each other via fast NVlink with a bandwidth of 600 GB/sec. Fast Mellanox HDR Infiniband switch enables communication among various nodes with an average bisectional bandwidth of $200$ Gb/sec/node. For our performance analysis, we employ a maximum of 512 GPUs residing in 64 nodes.

In this paper, we report the timings of our GPU-FFT. We start with the following real space field:
\bea
f(x,y,z) = 8\sin(x)\sin(2y)\sin(3z) \nonumber \\  
+ 8\sin(4x)\sin(5y) \sin(6z),\label{eq:function}
\eea
and perform a number of  forward and inverse FFT pairs.
As expected, for double precision, we recover the same function within an accuracy of $10^{-14}$ after the operation.  We performed FFTs on $1024^3$, $2048^3$, and $4096^3$ grids using 8-64, 8-128, and 64-512 GPUs respectively.  We averaged over 10 to 100 FFT pairs to reduce the errors in timings. We measured the timings for various operations in FFT using \texttt{MPI\_Wtime}. We focus on communication and computations times, which are denoted by $T_\mathrm{comp}$ and $T_\mathrm{comm}$ respectively. Note that the total time $T = T_\mathrm{comp}+ T_\mathrm{comm}$. The timings reported in this paper are for a pair of forward and inverse FFTs, and they are in milliseconds.

\begin{figure}[b!]
\begin{center}
\includegraphics[width=\linewidth]{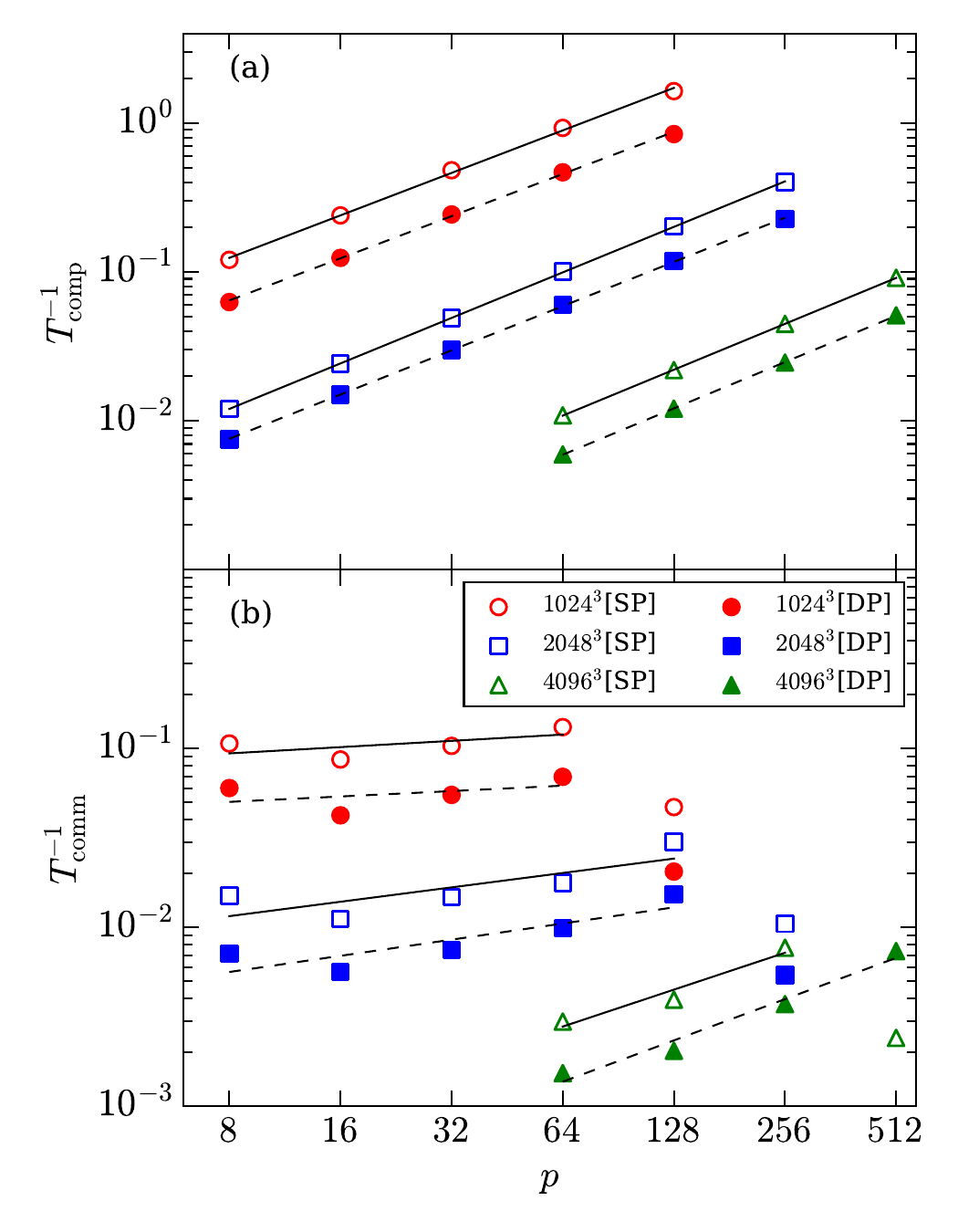}
\end{center}
\setlength{\abovecaptionskip}{0pt}
\caption{(a) Plots of inverse computation time, $T^{-1}_\mathrm{comp}$, vs.~$p$ (number of GPUs) for $1024^3$, $2048^3$, and $4096^3$ grids. (b) Plot of inverse communication time, $T^{-1}_\mathrm{comm}$, vs.~$p$. For all the figures, we employ filled symbols for double precision FFTs and unfilled ones for single precision FFTs.  Note that $T_\mathrm{comp}$ and $T_\mathrm{comm}$ (in milliseconds) are for a forward-inverse FFT pair.}
\label{fig:scaling_Selene_comp_comm}
\end{figure}

 In  Fig.~\ref{fig:scaling_Selene_comp_comm}(a,b), we exhibit respectively the inverses of the computation and communications times vs. $p$   for double-precision (DP) and single-precision (SP) FFTs on $1024^3$, $2048^3$ and $4096^3$ grids.  In this figure and in all subsequent ones, we exhibit the DP and SP timings using filled and unfilled symbols respectively. 

\begin{table*}[htbp]
\begin{center}
\caption{Scaling of our code on Selene: The exponents $\gamma_1$ for the computation time ($T_\mathrm{comp}$) scaling, $\gamma_2$ for the communication time  ($T_\mathrm{comm}$) scaling, and $\gamma$ for the total time  ($T$) scaling.  Here, we present single precision (SP) and double precision (DP) results.}
\hspace{20mm}
\begin{tabular}{c | c  c | c  c | c  c }
\hline \hline\\
Grid size &  \multicolumn{2}{c}{$\gamma_1$} &  \multicolumn{2}{c}{$\gamma_2$} &  \multicolumn{2}{c}{$\gamma$} \\
\hline
 & SP & DP & SP & DP & SP & DP\\
\hline
$1024^3$ & $0.98 \pm 0.01$  & $0.97 \pm 0.01$  &  $0.12 \pm 0.11$   &  $0.11 \pm 0.15$  &  $0.35 \pm 0.05$  &  $0.34 \pm 0.07$ \\
\\
$2048^3$ & $1.00 \pm 0.01$ & $1.00 \pm 0.002$  &  $0.27 \pm 0.12$  &  $0.30 \pm 0.10$  &  $0.49 \pm 0.06$  &  $0.48 \pm 0.05$\\
\\
$4096^3$ & $1.00 \pm 0.02$ & $1.00 \pm 0.02$ & $0.69 \pm 0.16$  & $0.64 \pm 0.13$ & $0.75 \pm 0.13$ & $0.71 \pm 0.11$\\
 \hline
\end{tabular}
\label{tab:fftk_exponents_selene}
\end{center}
\end{table*}
From the results exhibited in the figure, we observe that for all the three grids,  
\bea
	T_\mathrm{comp}  &=& C_1 p^{-\gamma_1}, \\
	T_\mathrm{comm}  &=& C_2 p^{-\gamma_2},\label{eq:scaling_relation}
\eea
where  $\gamma_{1}$, $\gamma_{2}$ are the exponents, and $C_{1}$, $C_{2}$ are constants, which were computed using regression analysis. Note, however,  that in Fig.~\ref{fig:scaling_Selene_comp_comm}(b), the last plot points for each grid, except for $4096^3$(DP), deviate from the best-fit curves. The exponents $\gamma_{1,2}$ are listed in Table~\ref{tab:fftk_exponents_selene}. The exponent for computation, $\gamma_1 \approx 1$ indicating that the inverse of the computation time scales linearly with the number of GPUs.  This is in expected lines.

However, the communication scaling is much inferior than the computation scaling due to the  extreme data transfers involved in FFT. The exponent $\gamma_2$ ranges from 0.12 to 0.75 with significant errors due to fluctuations, especially for $1024^3$ and $2048^3$ grid sizes. The poorest scaling is for $1024^3$ grid, while the reasonable scaling is observed for $4096^3$ grid. The inferior scaling for $1024^3$ grid is due to the fact that GPU-FFT is efficient within the DGX due to NVlink, but the performance drops suddenly when we go from 1 node to 2 and 4 nodes.  For $4096^3$ grid, the communication involves NVlink and Infiband for all the runs; hence, scaling is better for this grid. 

As shown in Fig.~\ref{fig:scaling_Selene_comp_comm}(b),  the  communication performance drops suddenly for the last plot points, except for $4096^3$(DP).  We observe that the chunk or packet size is 1 MB for the 5 plot points where the communication performance degrades considerably. The small packet size appears to be the reason for the drop in the performance of inter-GPU communication~\cite{MPI_Overheads_network:2006}. For large packets, GPUDirect RDMA and 3rd Gen NVLink are expected to provide efficient communication.

\begin{figure}[htbp]
\begin{center}
\includegraphics[width=\linewidth]{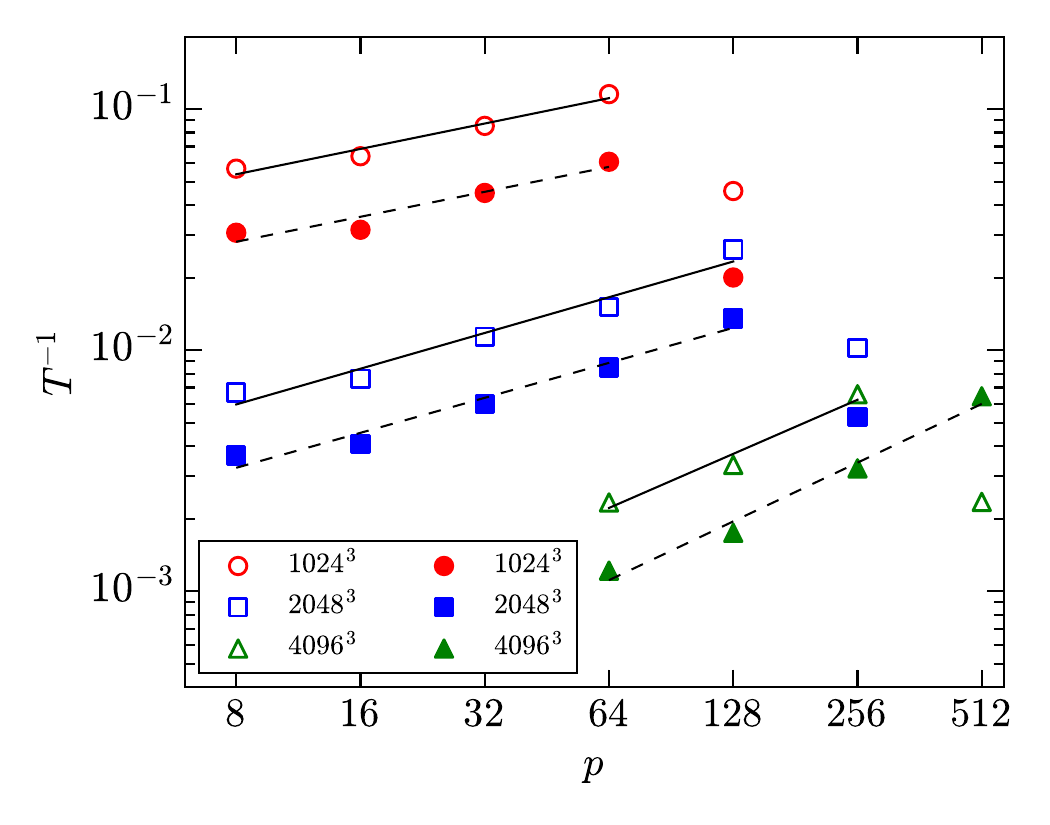}
\end{center}
\setlength{\abovecaptionskip}{0pt}
\caption{Plots illustrating the strong scaling ($T^{-1}$ vs.~$p$ ) for single and double precision FFTs of $1024^3$, $2048^3$, and $4096^3$ grids on {\em Selene}.  Note that $T$ (in millisecond) is for a forward-inverse FFT pairs.}
\label{fig:scaling_Selene_strong}
\end{figure}

Next, we report the strong and weak scaling of the total time. The strong scaling plots of  Fig.~\ref{fig:scaling_Selene_strong} reveal that
\be
 T = C p^{-\gamma}
 \ee
 with $\gamma$ ranging from 0.34 to 0.75 (see Table~\ref{tab:fftk_exponents_selene}). Clearly, the scaling is far from linear because of the expensive communication involved in FFT. 

\setlength{\tabcolsep}{10pt}
\begin{table}[h!]
\centering
\caption{Ratio of the communication time ($T_{\mathrm{comm}}$) and the total time ($T$), $r$, for various grid sizes with different number of GPUs ($p$) for single precision (SP) and double precision (DP) FFTs.}
\hspace{20mm}
\resizebox{\linewidth}{!}{
\begin{tabular}{c | c c | c c | c c}
\hline \hline \\[0.025 pt]
Grid Size & \multicolumn{2}{c}{$1024^3$} & \multicolumn{2}{c}{$2048^3$} & \multicolumn{2}{c}{$4096^3$} \\ \hline\\  [0.025 pt]

$p$    &    SP  &  DP  &  SP  &  DP & SP  &  DP \\ \hline

$8$  &  $0.53$   & $0.51$ &   $0.44$   & $0.51$ &  -  &  - \\[2 mm]

$16$ &  $0.74$   & $0.75$ &   $0.69$   & $0.73$ &   -  &  -   \\[2 mm]

$32$ &  $0.82$   & $0.82$ &  $0.77$    & $0.80$ & - & -   \\[2 mm]

$64$ &  $0.87$   & $0.87$  & $0.85$    &  $0.86$ & $0.79$  & $0.80$  \\[2 mm]

$128$ &  $0.97$   & $0.98$ &  $0.87$   &  $0.89$  &   $0.85$  & $0.86$    \\[2 mm]


$256$ &  -  &  - &  $0.97$   &  $0.98$  &   $0.85$  &  $0.87$  \\[2 mm]

$512$ &  -  &  - &  -   &  -  &   $0.97$  &  $0.87$  \\[2 mm]
\hline
\end{tabular}
}
\label{tab:comm/total-table}
\end{table}

\begin{figure}[b!]
\includegraphics[width=\linewidth]{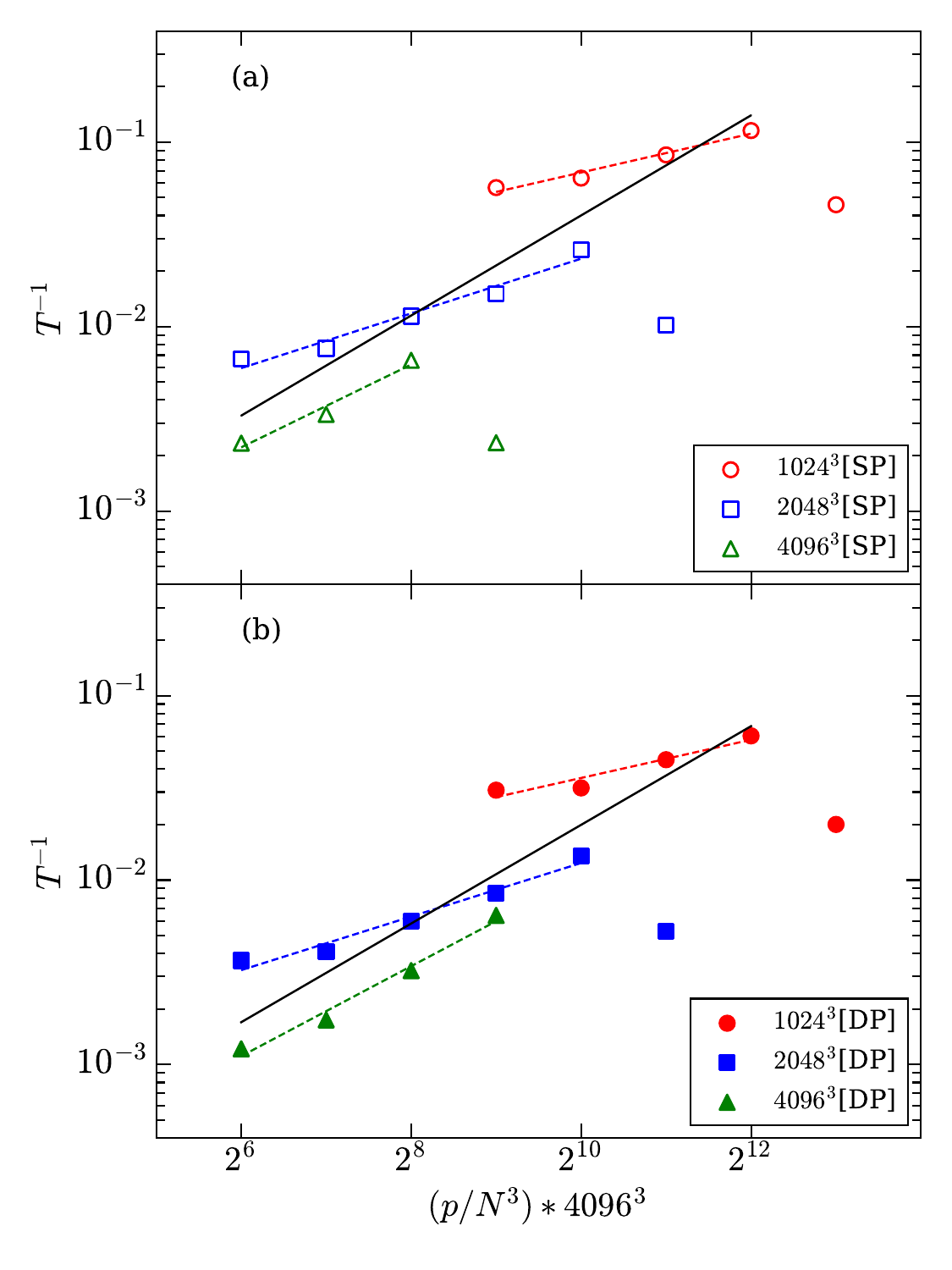}
\setlength{\abovecaptionskip}{0pt}
\caption{Plots illustrating the weak scaling  ($T^{-1}$ vs.~$(p/N^{3})*4096^{3}$) of our code  on  {\em Selene} for $1024^3$, $2048^3$, and $4096^3$ grids. }
\label{fig:weak_scaling_Selene}
\end{figure}

Now we present the weak scaling for all our runs. In  Fig.~\ref{fig:weak_scaling_Selene}(a,b), we plot $T^{-1}$ vs.~$(p/N^3)\times 4096^3$ for the single and double precision FFTs respectively. We observe that the collapse of the data points to a single curve is not very good. Rather, segments of the plots exhibit  behaviour similar to that for strong scaling (same local $\gamma$ as in Table~\ref{tab:fftk_exponents_selene} or Fig.~\ref{fig:scaling_Selene_comp_comm}(b)). An approximate fit to all the points yields weak-scaling exponent of approximately 0.9. Note, however, that this scaling exponent is not an accurate representation of the FFT performance.

\setlength{\tabcolsep}{10pt}
\begin{table}[h!]
\centering
\caption{TFLOPS rating of FFT implementation on  {\em Selene} using 8 to 512 GPUs. Note SP and DP stand for single and double precision respectively.}
\hspace{20mm}
\resizebox{\linewidth}{!}{
\begin{tabular}{c | c c | c c | c c}
\hline \hline \\
Grid Size & \multicolumn{2}{c}{$1024^3$} & \multicolumn{2}{c}{$2048^3$} & \multicolumn{2}{c}{$4096^3$} \\ \hline  
  
 $p$   &    SP  &  DP  &  SP  &  DP & SP  &  DP \\ \hline

$8$  &  $18$   & $10$ &   $19$   & $10$ &  -  &  - \\[2 mm]

$16$  &  $21$   & $10$ &   $22$   & $12$ &   -  &  -   \\[2 mm]

$32$  &  $27$   & $14$ &  $32$   & $17$ & - & -   \\[2 mm]

$64$  &  $37$   & $19$  & $43$   &  $24$ & $58$  & $30$  \\[2 mm]

$128$  &  $15$   & $6$  &  $74$   &  $38$  &   $83$  & $43$    \\[2 mm]

$256$  &  -   &  -  & $29$   &  $15$  &   $162$  &  $80$  \\[2 mm]

$512$  &  -  &  -  &  -  &  -  &  $58$   &  $159$ \\[2 mm]

\hline
\end{tabular}
}
\label{tab:tflops}
\end{table}

The strong and weak scaling plots indicate that communication dominates computation in GPU-FFT, as in other FFT applications~\cite{Pekurovsky:SIAM2012,Chatterjee:JPDC2018}. In Table~\ref{tab:comm/total-table}, we present the ratio of the communication time and the total time, $r$, for all our runs.  Note that the the fraction of time taken by computation is $1-r$. As shown in the table,  $r$ is minimum  for $2048^3$ grid with $p=8$. This is due to the fast communication via NVlink. The ratio $r$ increases for more GPUs due to internode communication via the Infiband switch. For the three grids, the highest values of $r$ are 0.98, 0.98, and 0.97 that occurs of packet size of 1 MB.  These values correspond to the last plot points of  Figs.~\ref{fig:scaling_Selene_comp_comm}, \ref{fig:scaling_Selene_strong}, and \ref{fig:weak_scaling_Selene} where the scaling deteriorates considerably. 

\begin{figure}[htbp!]
\includegraphics[width=\linewidth]{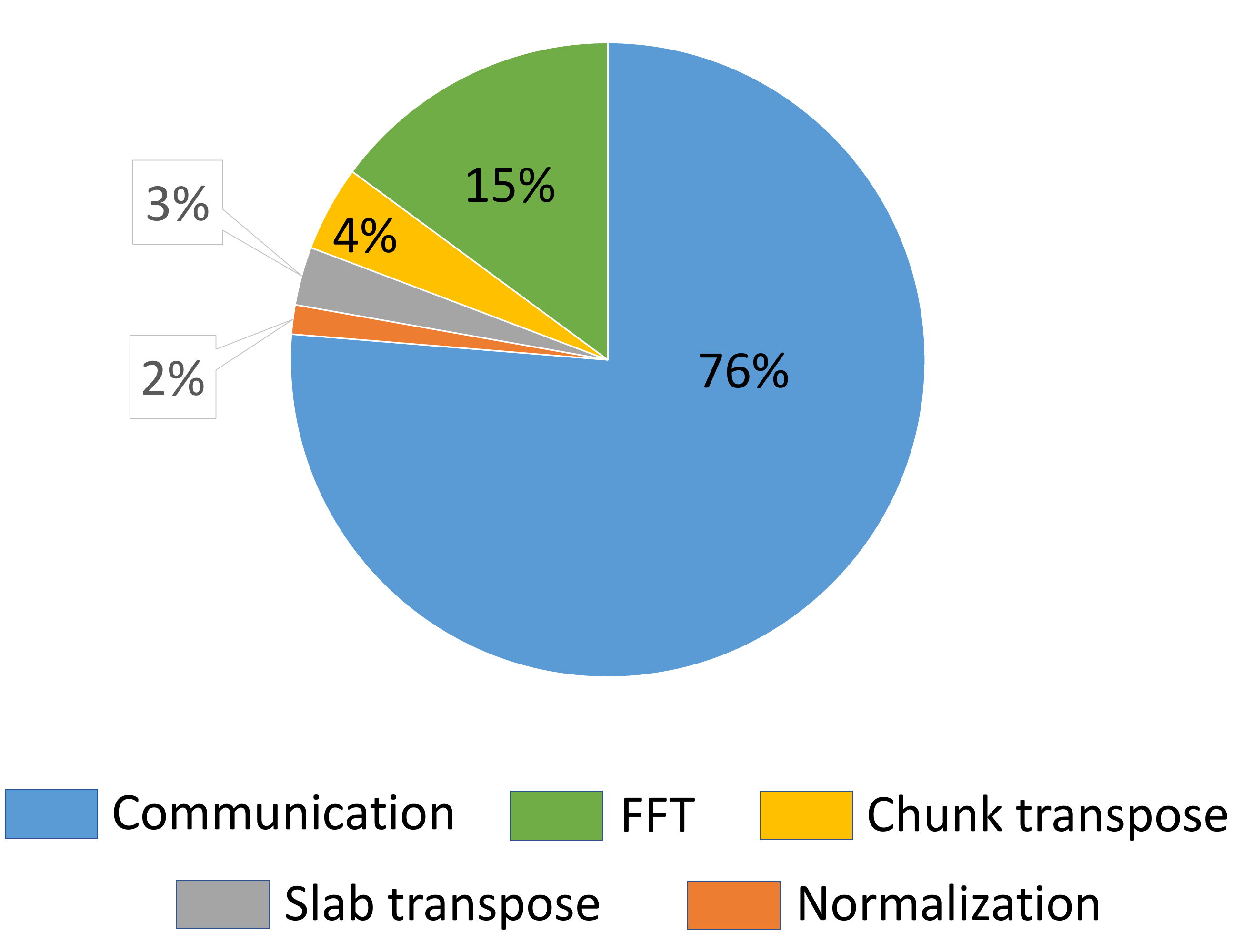}
\setlength{\abovecaptionskip}{0pt}
\caption{Distribution of timings of various functions during a FFT of $2048^3$ grid using 32 GPUs. Note that the distribution is strongly dependent on the number of GPUs and grid sizes.  }
\label{fig:pie_chart_time}
\end{figure}

In our paper, we report computation 
time as a sum of FFT, slab transpose, chunk transpose, and normalization. The communication time is the time spent on data transmission among the GPUs within and outside the node.  In Fig.~\ref{fig:pie_chart_time}, we illustrate the fraction of time taken by these functions for an FFT of $2048^3$ grid using 32 GPUs. Here, the most significant time,  76\%,  is taken by  communication. FFT takes 15\% of the total time, the slab and chunk transpose take respectively 3\% and 4\% of the total time, while normalization takes 2\% of the total time. Thus, the  computation takes 24\% of the total time, while communication takes 76\% of the total time.  We also observe that for an FFT of $2048^3$ grid with 128 GPUs, the  effective bandwidth across nodes is of 155 Gb/s, which is 75\% of the peak value. Hence, in Selene, the communication is quite efficient compared to the peak.  Unfortunately, the peak of communication  performance is much weaker that the peak of computation performance, which is the prime reason for degradation in the performance of parallel FFT.  We remark that the above ratios vary with the number of GPUs and grid size.

Finally, we present the FLOPS (floating point operations per second) rating of our GPU-FFT.  The total number of floating point operations for a pair of FFT is $2 \times 5N^{3}\mathrm{log}_{2}N^{3}$. Hence, the TFLOPS (tera floating point operations per second) rating of a GPU-FFT for a given grid is computed by dividing $10 N^{3}\mathrm{log}_{2}N^{3}\times 10^{-9}$ with the total time taken in seconds. In Table~\ref{tab:tflops} we list the TFLOPS ratings of the GPU-FFT for various grids and number of GPUs. For  single and double precision, the best  ratings are 162 TFLOPS and 159 TFLOPS  respectively, while the worst ratings are 15 TFLOPS and 7 TFLOPS respectively. Note that the single and double precision peak performances of a single A100 card are 19.5 TFLOPS and 9.7 TFLOPS respectively. Hence, the overall performance of our GPU-FFT is far less than the peak performance; this is due to the excessive communications involved in FFT.

\section{Comparison of GPU-FFT with other FFTs}
\label{sec:comparisons}

First, we compare the performance of our GPU-FFT with FFTK, which is a multicore FFT library. Chatterjee {\em et al.}~\cite{Chatterjee:JPDC2018} ran FFTK on {\em Shaheen II} (Cray XC40) for  $1536^3$ and $3072^3$ grids using a maximum of 196608 cores (of Intel Haswell). They observed that FFTK scaled well up to 196608 processors of Cray XC40 with $\gamma_1 \approx 1$,  $\gamma_2$ taking values between 0.43 to 0.60 depending on the grid size, and  $\gamma \approx \gamma_2$. In comparison, on {\em Selene}, our GPU-FFT yields $\gamma_1 \approx 1$ for all the runs, but $\gamma_2$ and $\gamma$ are near FFTK's values only for the largest grid ($4096^3$). Refer to our earlier discussion why $\gamma_2$ and $\gamma$ are small for smaller grids.
\setlength{\tabcolsep}{10pt}
\begin{table*}[t!]
\centering
\caption{A comparison between the performance of cuFFTXT and our GPU-FFT:  Total times (in milliseconds) for  a pair of single precision (SP) and double precision (DP) forward and backward FFTs of different grid sizes using  2 to 8 GPUs.}
\hspace{20mm}
\resizebox{\linewidth}{!}{
\begin{tabular}{c | c c c c| c c c c| c c c c}
\hline
Grid Size &  \multicolumn{4}{c}{$512^3$} & \multicolumn{4}{c}{$1024^3$} &  \multicolumn{4}{c}{$2048^3$} \\ \hline

       &    \multicolumn{2}{c}{cuFFTXT}  &    \multicolumn{2}{c}{GPU-FFT}  &    \multicolumn{2}{c}{cuFFTXT}  &    \multicolumn{2}{c}{GPU-FFT}  &  \multicolumn{2}{c}{cuFFTXT}  &  \multicolumn{2}{c}{GPU-FFT} \\ \hline

$p$    &    SP  &  DP  &    SP  &  DP  &    SP  &  DP  &    SP  &  DP  &  SP  &  DP  &  SP  &  DP \\ \hline

$2$    &  $3.6$  &  $6.5$  &  $5.6$  &  -  &  $27$  &  $54$  &  $44$  &  -  &  -  &  -  &  -  &  - \\[2 mm]

$4$    &  $2.3$  &  $4.3$  &  $3.6$  &  -  &  $15$  &  $30$  &  $28$  &  -  &  -  &  -  &  -  &  - \\[2 mm]

$8$    &  $1.7$  &  $5.1$  &  $2.2$  &  -  &  $11$  &  $19$  &  $17.6$  &  $32$  &  $76$  &  $132$  &  $150$  &  $273$  \\[2 mm]

\hline
\end{tabular}
}
\label{tab:cufftxt:selene:MPI}
\end{table*}

Let us compare the overall performance of GPU-FFT and FFTK on  {\em Selene} and {\em Shaheen II} respectively. We digitized Fig. 6 of Chatterjee {\em et al.}~\cite{Chatterjee:JPDC2018} and observed that  a pair of double-precision forward and backward FFTs of $1536^3$ and $3072^{3}$ grids using 196608 processors took  42 and 179 milliseconds respectively. In comparison, on {\em Selene}, single-precision and double-precision GPU-FFTs of $2048^3$ grid using 128 GPUs took 38 milliseconds and {\color{blue}74} milliseconds respectively. Thus, 128 GPUs yield performance comparable to 196608 Intel Haswell cores (or 12288 processors) of Cray XC40.  The improved performance on {\em Selene} is due to the fast computation by A100 GPUs and efficient inter-GPU communication via NVlink.  These observations indicate that GPUs may provide cost-effective solutions to extreme HPC problems. It will be interesting to compare the performance of multicore and GPU FFTs on other platforms. Also, it is important to compare the ratio of communication time and the total time for the multicore and GPU FFTs.

There are other GPU-FFT libraries, e.g., heFFTe, cuFFTXT, AccFFT, etc.~\cite{heffte_2020,acfft:2015}. Note that most of these libraries operate within a node due to a lack of multinode support.  Recently, NVIDIA has announced a multi-node GPU-FFT library called cuFFTMp for early access.  Due to lack of data, we are not able to compare the performance of our code with that of cuFFTMp.  Here, we compare our code performance with those of heFFTe and cuFFTXT within a node.

We compare the timings of cuFFTXT and our GPU-FFT library on a single DGX box for $512^{3}$, $1024^{3}$, and $2048^{3}$ grids. See  Table.~\ref{tab:cufftxt:selene:MPI} for details. The performance of our FFT is approximately half of that of cuFFTXT. The reason for the drop in the performance of our GPU-FFT may be due to MPI overheads.   We plan to work on detailed investigation of this aspect in future. Regarding the heFFTe library, the timings of GPU-FFT of  $1024^{3}$ grid using a box with 6 V100 GPUs is approximately 500 milliseconds~\cite{heffte_2020}.  In comparison, our GPU-FFT library takes 30 milliseconds on 4 A100 cards.  Note, however, that a V100 card is less efficient than A100 card.

Ravikumar {\em et al.}~\cite{young_PK:ACM2019} performed spectral simulation of turbulent flows using their own  asynchronous batched  GPU-FFT.  In addition,  Bak {\em et al.}~\cite{Bak:PC2022} performed weak scaling of their synchronous non-batched GPU-FFT on  {\em Summit}. They observed that a FFT of $3072^3$ grid using 96 V100 GPUs (16 nodes) of Summit took $550$ milliseconds~\cite{comm}.  In comparison, on {\em Selene}, our GPU-FFT of $4096^3$ grid using 128 A100 GPUs took approximately 287 milliseconds. Hence, the performance of the two GPU-FFTs are comparable.   It is likely that our strategy of avoidance of device-to-host communication during \texttt{Alltoall} may yield better efficiency. 

\section{Conclusions}
\label{sec:conclusions}

Present  GPUs are  computationally efficient. Within a node,  NVlink provides  fast inter-GPU communication with a  bandwidth of 600 GB/sec. Due to these reasons, a large number of current HPC systems contain GPUs as accelerators. Encouraged by availability of such efficient systems, we created a multi-node GPU-FFT  on CUDA platform and tested them on {\em Selene} supercomputer.

In this paper, we present the algorithm of our multi-node GPU-FFT. We employ a maximum of 512 GPUs for our runs, which is lower than the  grid size along any direction ($N$ of $N^3$). Hence, we employ slab decomposition for the data division. We employ MPI for communication among the GPUs within a node and across nodes. We tested our code  using a maximum of 64 DGX boxes or 512 GPUs of {\em Selene}. A summary of our results are as follow:

\begin{enumerate}
    \item As expected, the computation component (FFT + slab and chunk transpose) scales linearly with the number of GPUs.
    
    \item The scaling of inter-GPU communication depends critically on the grid size and number of GPUs. The communication is very efficient within a node due to NVlink.  But, the communication speed drops significantly when we go from one node to two or higher number of nodes. For efficient communication, we need large grid ($4096^3$) with proportionally large number of GPUs. The scaling drops when the packet size comes down to 1 MB or less.
    
    \item Due to the computation efficiency of A100 cads and efficient communication via NVlink, the execution time for $2048^3$ grid with 128 GPU cards is comparable to the timings for $1536^3$ grid using 12288 Haswell processors.
    
\end{enumerate}
Based on these results, we deduce that the best performance on a GPU cluster is realized either within a single DGX box for a relative small grid, or with many more GPUs for a large grid. The computation performance drops significantly when we go from one DGX box to two boxes. 

In summary, multi-node GPU-FFT provides an excellent opportunity to solve large-scale spectral problems. Our GPU-FFT library is an open-source library (in contrast to cuFFTMp and Ravikumar {\em et al.}'s FFT~\cite{young_PK:ACM2019}), hence it will be useful to community for experimentation. 

\section{Acknowledgement}
\label{sec:acknowledgement}
We thank Kiran Ravikumar, P. K. Yeung, Manish Modani, Isaac George, Jayesh Badwaik, Preeti Malakar, and Anando  Chatterjee for useful discussions. We thank Nvidia for an access to {\em Selene}, where scaling studies were performed, and to CDAC for an access to  {\em PARAM Siddhi-AI}, where initial testing of our GPU-FFT was done.  Soumyadeep Chatterjee is supported by INSPIRE fellowship (IF180094) of Department of Science \& Technology, India.



\end{document}